\documentclass[fleqn,usenatbib]{mnras}

\makeatletter
\renewcommand{\@journal}{}
\makeatother

\usepackage{newtxtext,newtxmath}

\usepackage[T1]{fontenc}

\DeclareRobustCommand{\VAN}[3]{#2}
\let\VANthebibliography\thebibliography
\def\thebibliography{\DeclareRobustCommand{\VAN}[3]{##3}\VANthebibliography}

\usepackage{amsmath,mathrsfs,graphicx,adjustbox,bm,graphicx}    

\def\mnras{Mon. Not. R. Astron. Soc. }

\def\apj{Astrophys. J.}
\def\apjl{Astrophys. J. Lett.}
\def\apjs{Astrophys. J., Suppl. Ser.}
\def\prd{Phys. Rev. D}

\def\aap{Astron. Astrophys.}

\usepackage{lineno}

\DeclareFontFamily{U}{wncy}{}
\DeclareFontShape{U}{wncy}{m}{n}{<->wncyr10}{}
\DeclareSymbolFont{mcy}{U}{wncy}{m}{n}
\DeclareMathSymbol{\Sh}{\mathord}{mcy}{"58}

\title{Popcorn EMRIs: Transient Gravitational Wave Signals and Their Analysis in Schwartz Space}

\author[Pau Amaro Seoane, Kostas Tzanavaris]{Pau Amaro Seoane$^{1,2,3,4}$\thanks{E-mail: amaro@upv.es},
Kostas Tzanavaris,$^{4,5,6}$\thanks{E-mail: konstantinos.tzanavaris@aei.mpg.de}
\\
$^{1}$Universitat Polit\`ecnica de Val\`encia, Spain\\
$^{2}$Max Planck Institute for Extraterrestrial Physics, Garching, Germany\\
$^{3}$The Kavli Institute for Astronomy and Astrophysics, PKU, Beijing, China\\
$^{4}$Higgs Centre for Theoretical Physics, James Clerk Maxwell Building, Edinburgh EH9 3FD, UK\\
$^{5}$Max Planck Institute for Gravitational Physics (Albert Einstein Institute), D-30167 Hannover, Germany\\
$^{6}$Leibniz Universit{\"a}t Hannover, Hannover, D-30167, Germany
}

\date{}

\pubyear{\the\year{}}

\makeatletter
\def\ps@headings{%
  \let\@mkboth\markboth
  \def\@oddhead{\Large\hfill\it\@shorttitle
    \hspace{1.5em}\rm\thepage}%
  \def\@evenhead{\Large\thepage
    \hspace{1.5em}\it\@shortauthor\hfill}%
  \def\@oddfoot{}%
  \def\@evenfoot{}%
}
\makeatother
\begin{document}
\label{firstpage}
\pagerange{\pageref{firstpage}--\pageref{lastpage}}
\maketitle

\begin{abstract}
This study investigates extreme-mass ratio inspirals (EMRIs) with orbital periods exceeding the observation windows of mHz gravitational wave detectors. During their early, highly eccentric stages, these systems produce periodic gravitational wave eruptions at pericentre, separated by long quiescent periods. We define these systems as ``popcorn EMRIs''—bound, coherent binaries in a deterministic evolutionary phase. This distinguishes them from stochastic, non-coherent bursts associated with parabolic fly-bys. Using a steady-state model based on the phase-space continuity equation, we estimate the population within a Milky Way-like galaxy. Because these estimates depend on sensitive astrophysical variables like stellar relaxation and density profiles, our figures are illustrative rather than absolute. For a one-year observation baseline, we estimate a steady-state population of 200 to 2,000 popcorn EMRIs, significantly outnumbering the 10 to 88 continuously observable EMRIs. This yields an expected rate of 5 to 44 detectable eruptions per year. Despite high signal-to-noise ratios (exceeding 700 for Galactic Centre events), the low duty cycle confirms their transient nature. Standard data processing techniques, designed for continuous signals, often employ windowing that distorts eruption morphology and artificially reduces the signal-to-noise ratio. To address this, we establish a mathematical apparatus using based on standard smoothing techniques, in the theory of tempered distributions. We demonstrate that standard windowing techniques irreversibly degrade the signal-to-noise ratio of transient sources when the taper overlaps the eruption, and we utilize Schwartz space regularization to justify the direct Fourier transform as the energy-preserving estimator.
\end{abstract}




\section{Introduction}

The inspiral of a stellar-mass compact object into a massive black hole (MBH), known as an extreme-mass ratio inspiral (EMRI), represents a critical target for the Laser Interferometer Space Antenna (LISA). These systems provide unparalleled opportunities to probe the geometry of spacetime in the strong-field regime \citep{AS2007,BabakEtAl2017,AS2018,AS2022}.

The evolution of EMRIs is dominated by the emission of gravitational waves (GWs). The timescale for this process scales strongly with the semi-major axis ($T_{\rm GW} \propto a^4$) \citep{Peters1964}, implying that EMRIs spend the vast majority of their lifetime in the early stages of inspiral, characterized by wide orbits and long orbital periods $P$. Although the merger rate at the Galactic Centre (GC) is relatively low, the extended duration of the early inspiral phase suggests a substantial steady-state population of Early EMRIs (E-EMRIs) may currently reside within the LISA frequency band \citep{amaro2024mono}.

This work focuses on the subset of E-EMRIs whose orbital periods exceed the duration of observation, $P > T_{\rm obs}$. We adopt a conservative baseline of $T_{\rm obs}=1$ year for primary analysis, while also considering the nominal 4-year mission duration. EMRIs are generally formed with very high eccentricities \citep{amaro2019extremely}. In such configurations, the GW emission is concentrated into brief, intense eruptions during the pericentre passage, which has a duration $dt_{\rm peri}$. When $dt_{\rm peri} \ll T_{\rm obs} < P$, the observable signal consists of isolated transient events separated by extended periods of quiescence. We designate these discontinuous sources as ``popcorn EMRIs''.

The study of gravitational wave bursts from galactic nuclei has a rich history. \cite{RubboEtAl2006} estimated rates for extreme mass ratio bursts (EMRBs) for LISA, whilst \cite{ToonenEtAl2009} and \cite{HopmanEtAl2007} investigated the background and rates from star-MBH fly-bys, noting the sensitivity to mass segregation and a realistic inner density profile. \cite{BerryGair2013} and \cite{BerryGair2013b} provided detailed analyses of parameter estimation from such bursts, and \cite{FanEtAl2022} explored detection with TianQin. We adopt the term ``popcorn EMRIs'' to distinguish these bound sources from parabolic fly-bys. (Recurring eruptions from bound, highly eccentric orbits have previously been studied by \cite{OliverEtAl2024} and \cite{OliverEtAl2026} using numerical kludge waveforms.) They are coherent sources evolving deterministically under radiation reaction, linking the population directly to the steady-state EMRI channel. We therefore term the signal from a single pericentre passage of a popcorn EMRI an ``eruption'' to distinguish it from stochastic bursts.

We estimate the population and detection rate of popcorn EMRIs at the GC using steady-state distribution models derived from the continuity equation in phase space. These estimates rely on analytical models that necessarily simplify the complex dynamics of the Galactic Centre; therefore, the derived numbers must be interpreted as illustrative rather than definitive. We emphasize that these are based on the analytical estimates of \cite{amaro2024mono}, which need to be improved. These are not absolute numbers. We subsequently analyze their observational signatures, calculating the expected eruption rate, duty cycle, and signal-to-noise ratios.

The spectral analysis of these transient signals presents significant challenges. Conventional Fourier techniques are optimized for continuous signals. When analyzing finite segments of continuous data, standard practice involves the application of window functions (tapering) to mitigate spectral leakage. However, popcorn EMRIs are intrinsically transient. Applying windowing techniques to these eruptions distorts the physical morphology of the signal and can lead to substantial underestimation of the signal energy and SNR.

To address this fundamental issue, we develop a mathematical apparatus using based on standard smoothing techniques, in the theory of tempered distributions.\footnote{See \cite{Hormander1990ALPDO1,Hormander2005ALPDO2} for a review.} This formalism provides a mathematically sound definition for the Fourier transform of transient and potentially discontinuous eruptions. It circumvents the artifacts associated with ad hoc windowing and justifies the application of the direct Fourier transform for the spectral analysis of these signals. This analysis is required because, unlike continuous signals where windowing removes boundary artifacts, for a contained transient, windowing removes physical signal energy. 

Convolving the signal in time with a mollifier corresponds to multiplying the Fourier transform by a Gaussian window in frequency. This mechanism provides an alternative to standard data analysis, which applies tapers in the time domain. Because eruptions have short durations, tapering the time series forces the physical signal to zero at the observation boundaries, which decreases the signal-to-noise ratio and dissipates energy. The mollifier operates as a smooth low-pass filter that prevents Gibbs ringing and retains the core energy. The convolution recovers the unattenuated signal power in the asymptotic limit.

Evaluating the limit as $n$ approaches infinity flattens the frequency window to unity, proving that the untapered Fourier transform maintains the total signal energy. This limit illustrates the Heisenberg-Gabor limit. The time-domain mollifier width scales proportionally to $1/\sqrt{n}$ and converges to a Dirac delta distribution. In the frequency domain, the width increases proportionally to $\sqrt{n}$. Measuring the onset and cessation of these eruptions requires a large frequency range, because the signals distribute power across a wide spectrum. Standard data analysis windowing removes this necessary spectral distribution, while the convolution method preserves it.

\section{Astrophysical Estimation of the Popcorn EMRI Population}

We derive an estimate for the steady-state population of popcorn EMRIs in a Milky Way-like galaxy based on the analytical astrophysical model presented in \cite{amaro2024mono}. This calculation relies on necessary simplifications regarding the stellar distribution and mass function. We utilize an illustrative two-component model ($10\,M_{\odot}$ and $40\,M_{\odot}$). We perform the analysis for two observational timescales: a conservative $T_{\rm obs}=1$ year and the nominal $T_{\rm obs}=4$ years.

\subsection{Model Parameters and Definitions}

We consider a central MBH of mass $M_{\rm BH}=4.3\times 10^{6}M_{\odot}$. We adopt an influence radius $R_h=3$ pc, consistent with current observations of the Galactic Centre \citep{GenzelEtAl10}. The population model {adopted from \cite{amaro2024mono}} divides EMRIs into three regimes based on the semi-major axis $a$: $N_I$ (circular, $a<a_{\rm thr}$), $N_{II}$ (eccentric, detectable, $a_{\rm thr}<a<a_{\rm band}$), and $N_{III}$ (eccentric, early stage, $a_{\rm band}<a<a_{\rm crit}$).

These radii are defined in \cite{amaro2019extremely}. This work identifies $a_{\rm crit}$ as the semi-major axis where the threshold curve separating the relaxation and gravitational wave regimes merges with the last stable orbit. We express this critical radius as
\begin{equation}
    a_{\rm crit} = \epsilon \, R_{0} \left[ {\cal W}(\iota, s)^{5/2} N_{0} \ln(\Lambda) \left( \frac{M_{\rm BH}}{m_{\rm bh}} \right)^{-1} \right]^{1/(\gamma-3)},
    \label{eq:a_crit}
\end{equation}

\noindent where $\epsilon$ follows a function of the power-law index $\gamma$, $M_{\rm BH}$ is the mass of the supermassive black hole and $m_{\rm bh}$ the mass of the stellar-mass black hole. 

The function ${\cal W}(\iota, s)$ accounts for the impact of the asymmetry between prograde and retrograde orbits on the location of the last stable orbit for a kerr massive black hole \citep{amaro2013role}. This function depends on the inclination of the orbit $\iota$ and the magnitude of the spin of the central massive black hole $s$. By incorporating this term, we take into account the pericentre distance of the last stable parabolic orbit relative to the standard Schwarzschild case. 

The quantity $a_{\rm band}$ marks the semi-major axis at which the gravitational waves reach frequencies detectable by the LISA observatory. The same work defines $a_{\rm break}$ as the scale where the orbit transitions from high eccentricity to a circular state. finally, the minimum semi-major axis $a_{\rm min}$ represents the distance containing at least one object to initiate the integration, which typically equates to the plunge radius.

The steady-state distribution of EMRIs is derived from the continuity equation in the phase space of semi-major axis $a$ \citep{amaro2024mono}. In the eccentric regimes ($N_{II}$ and $N_{III}$), {where the evolution is dominated by gravitational wave emission,} this analysis yields $dN/da \propto a^{-1/2}$. This scaling arises because the steady-state condition implies a constant flux of objects through energy space, $\dot{a} (dN/da) = \text{constant}$. For highly eccentric orbits near the loss cone, the time-averaged evolution rate scales as $\dot{a} \propto a^{1/2}$ within the model's approximations \citep{amaro2019extremely}. This analytical result is consistent with findings from numerical simulations \citep{ZhangAmaroSeoane2025}. Consequently, the number of sources $N(a_1, a_2)$ between semi-major axes $a_1$ and $a_2$ is:
\begin{equation}
    \label{eq:N_scaling_a}
    N(a_1, a_2) \propto \int_{a_1}^{a_2} a^{-1/2} da = 2(a_2^{1/2} - a_1^{1/2}).
\end{equation}

\noindent The analysis relies on the population estimates derived in \cite{amaro2024mono}, which incorporate the effects of MBH spin ($s$) and orbital inclination ($\iota$) via the function ${\cal W}(\iota,\,{\rm s})$ \citep{amaro2013role}. {These effects shape the normalization and boundaries of the steady-state populations $N_{I, II, III}(s, \iota)$.}

\subsection{Orbital Period Thresholds}

We define the threshold for popcorn EMRIs by $P > T_{\rm obs}$. We calculate the corresponding semi-major axes $a_{T}$ using Kepler's third law.
\begin{align}
    a_{T, 1\text{yr}} &\approx 7.884\times 10^{-4}\,\text{pc}, \\
    a_{T, 4\text{yr}} &\approx 1.987\times 10^{-3}\,\text{pc}.
\end{align}

\noindent Sources with $a>a_{T}$ are popcorn EMRIs ($N_{\rm pop}$), and sources with $a<a_{T}$ are continuous ($N_{\rm mono}$).

\subsection{Population Analysis for $10 M_\odot$ Compact Objects}

We utilize boundaries from \cite{amaro2024mono}: $a_{\rm thr} = 2.17\times 10^{-6}$ pc and $a_{\rm band} = 4.52\times 10^{-3}$ pc. We observe $a_{\rm thr} < a_{T} < a_{\rm band}$ for both timescales. In region $N_{III}$ ($a>a_{\rm band}$), all sources are popcorn EMRIs. In region $N_{II}$ ($a_{\rm thr}<a<a_{\rm band}$), the population is split by $a_{T}$. The fraction of popcorn EMRIs $F_{II, \rm pop}$ is:
\begin{align}
    F_{II, \rm pop}(T_{\rm obs}) &= \frac{a_{\rm band}^{1/2} - a_{T}^{1/2}}{a_{\rm band}^{1/2} - a_{\rm thr}^{1/2}}.
\end{align}

\noindent $F_{II, \rm pop}(1\text{yr}) \approx 0.5954$; $F_{II, \rm pop}(4\text{yr}) \approx 0.3445$.

We utilize the conservative ranges for $R_h=3$ pc: $N_{II} \in [0.05, 0.45]$ and $N_{III} \in [1, 8]$. We evaluate the extrema of $N_{\rm Pop, 10}(s, \iota) = N_{III}(s, \iota) + F_{II, \rm pop} N_{II}(s, \iota)$, accounting for the anti-correlation between $N_{II}$ and $N_{III}$. For $T_{\rm obs}=1$ year,
\begin{align*}
    N_{\rm Pop, 10}^{\rm max} &\approx 8 + 0.5954 \times 0.05 \approx 8.030, \\
    N_{\rm Pop, 10}^{\rm min} &\approx 1 + 0.5954 \times 0.45 \approx 1.268.
\end{align*}

\noindent For $T_{\rm obs}=4$ years the results are not very different,
\begin{align*}
    N_{\rm Pop, 10}^{\rm max} &\approx 8 + 0.3445 \times 0.05 \approx 8.017, \\
    N_{\rm Pop, 10}^{\rm min} &\approx 1 + 0.3445 \times 0.45 \approx 1.155.
\end{align*}

\subsection{Population Analysis for $40 M_\odot$ Compact Objects}

We utilize boundaries from \cite{amaro2024mono}: $a_{\rm thr} = 5.59\times 10^{-6}$ pc and $a_{\rm band} = 2.45\times 10^{-4}$ pc. We observe $a_{\rm band} < a_{T}$ for both timescales.

In region $N_{II}$, all sources are continuous. In region $N_{III}$ ($a>a_{\rm band}$), the population is split by $a_{T}$. The continuous population is $N_{III, \rm cont} = N(a_{\rm band}, a_{T})$. We use the steady-state ratio:
\begin{equation}
    R_{T}(T_{\rm obs}) = \frac{N_{III, \rm cont}}{N_{II}} = \frac{a_{T}^{1/2} - a_{\rm band}^{1/2}}{a_{\rm band}^{1/2} - a_{\rm thr}^{1/2}}.
\end{equation}

\noindent $R_{T}(1\text{yr}) \approx 0.9351$; $R_{T}(4\text{yr}) \approx 2.177$. The total popcorn population is $N_{\rm Pop, 40}(s, \iota) = N_{III}(s, \iota) - R_{T} N_{II}(s, \iota)$. We utilize the conservative ranges: $N_{II} \in [5, 45]$ and $N_{III} \in [250, 2000]$. Hence, for $T_{\rm obs}=1$ year:
\begin{align*}
    N_{\rm Pop, 40}^{\rm max} &\approx 2000 - 0.9351 \times 5 \approx 1995.3, \\
    N_{\rm Pop, 40}^{\rm min} &\approx 250 - 0.9351 \times 45 \approx 207.9.
\end{align*}

\noindent For $T_{\rm obs}=4$ years the results are again similar,
\begin{align*}
    N_{\rm Pop, 40}^{\rm max} &\approx 2000 - 2.177 \times 5 \approx 1989.1, \\
    N_{\rm Pop, 40}^{\rm min} &\approx 250 - 2.177 \times 45 \approx 152.0.
\end{align*}

\subsection{Total Estimated Population and Limitations}

We sum the contributions from the two illustrative masses, $N_{\rm Pop, Total} = N_{\rm Pop, 10} + N_{\rm Pop, 40}$, aligning the corresponding extrema. For $T_{\rm obs}=1$ year: $N_{\rm Pop, Total} \approx 209$ to $2003$.
For $T_{\rm obs}=4$ years: $N_{\rm Pop, Total} \approx 153$ to $1997$.

We also calculate the total continuous population $N_{\rm mono}$ by summing the continuous components ($N_I$, $N_{II, \rm cont}$, $N_{III, \rm cont}$) across both mass bins, utilizing the $N_I$ ranges from the underlying model. For $T_{\rm obs}=1$ year: $N_{\rm mono} \approx 10$ to $88$. For $T_{\rm obs}=4$ years: $N_{\rm mono} \approx 16$ to $144$.

The population estimates derived here must be regarded as purely illustrative due to the inherent limitations of the underlying analytical steady-state model. As emphasized, this is just an illustration based on the analytical estimates of Amaro Seoane, which need to be improved. These are not absolute numbers. The derivation relies on solving the continuity equation under several idealized assumptions. The normalization of the steady-state population $N$ is proportional to the merger rate $\Gamma$, which depends strongly on the assumed stellar density profile $\rho(r) \propto r^{-\gamma}$ and the total number of compact objects $N_0$. Specifically, the analytical derivation of $\Gamma$ \citep[see][]{amaro2019extremely} shows a complex dependence on these poorly constrained parameters. Furthermore, the model assumes a sharp transition at the critical radius $a_{\rm crit}$ between relaxation-dominated and GW-dominated regimes. This transition is modeled by matching the fluxes, $\mathcal{F}_{\rm GW} = \mathcal{F}_{\rm relax}$, ignoring the interplay between these processes in the transition region. The model also employs a simplified two-component mass function, $f(m) = c_1 \delta(m-10M_\odot) + c_2 \delta(m-40M_\odot)$. The scaling $dN/da \propto a^{-1/2}$ is specific to the high-eccentricity limit ($e\to 1$) and the adopted relaxation model. Deviations from these idealized conditions, such as anisotropies, are not captured, potentially leading to significant systematic uncertainties in the absolute number of sources.

\section{Observational Signatures: Popcorn Rate and Detectability}

We analyze the observational characteristics of the E-EMRI population, utilizing the astrophysical estimates derived above and the scaling laws inherent to the steady-state, GW-dominated evolution.

\subsection{Population Model and Scaling Laws}

The steady-state distribution in the eccentric, GW-dominated regime, $dN/da \propto a^{-1/2}$, implies a distribution of periods (using $P \propto a^{3/2}$):
\begin{equation}
    \label{eq:period_distribution_model_sec3}
    \frac{dN}{dP} = C P^{-2/3}.
\end{equation}

\noindent Assuming $P_{\rm min} \ll T_{\rm obs} \ll P_{\rm max}$, the continuous population is
\begin{equation}
    N_{\rm mono}(T_{\rm obs}) = \int_{P_{\rm min}}^{T_{\rm obs}} C P^{-2/3} dP \approx 3 C T_{\rm obs}^{1/3}.
\end{equation}

\subsection{Expected Rate}

The expected number of popcorn eruptions $N_{\rm erupt}$ observed during $T_{\rm obs}$ from the population ($P > T_{\rm obs}$) is:
\begin{align}
    N_{\rm erupt}(T_{\rm obs}) &= \int_{T_{\rm obs}}^{P_{\rm max}} \frac{T_{\rm obs}}{P} \frac{dN}{dP} dP \approx \frac{3}{2} C T_{\rm obs}^{1/3}.
\end{align}

\noindent The ratio of expected eruptions to continuous sources is independent of $T_{\rm obs}$ under these assumptions:
\begin{equation}
    \frac{N_{\rm erupt}}{N_{\rm mono}} \approx 1/2.
\end{equation}

\noindent The average eruption rate is $\mathcal{R} = N_{\rm erupt}/T_{\rm obs} \propto T_{\rm obs}^{-2/3}$. Using the derived ranges for $N_{\rm mono}$, for $T_{\rm obs}=1$ year ($N_{\rm mono} \in [10, 88]$), $\mathcal{R} \approx 5$ to $44$ per year. For $T_{\rm obs}=4$ years ($N_{\rm mono} \in [16, 144]$): $\mathcal{R} \approx 2$ to $18$ per year.

\subsection{Duty Cycle}

We estimate the duty cycle $D = \mathcal{R} \langle \tau_{\text{obs}} \rangle$, where $\mathcal{R}$ is the event rate and $\langle \tau_{\text{obs}} \rangle$ is the observable duration of the signal.

It is necessary to distinguish between the dynamical timescale of the pericentre passage and the effective observational duration. The dynamical timescale, characterizing the signal peak, is $\tau_{\text{dyn}} \approx \sqrt{R_p^3 / (2GM)} \approx 474 \text{ s}$.

However, the signal remains detectable as long as its instantaneous frequency exceeds the detector's low-frequency noise floor. For the illustrative system ($P = 1 \text{ year}$, $e \approx 0.9974$) orbiting Sgr A*, the gravitational wave frequency scales as $f(t) \propto r(t)^{-3/2}$. Assuming a conservative observational floor of $f_{\min} \approx 10^{-4} \text{ Hz}$, the source remains "in-band" while $r(t) \lesssim 2.8 R_p$. Solving Kepler's equation for this condition yields an observable duration of:
\begin{equation}
\langle \tau_{\text{obs}} \rangle \approx 4100 \text{ s}.
\end{equation}

\noindent This is approximately an order of magnitude longer than the dynamical peak width.
Consequently, the estimated duty cycles increase for $T_{\text{obs}} = 1 \text{ year}$, $D \approx 6.5 \times 10^{-4} - 5.7 \times 10^{-3}$. For $T_{\text{obs}} = 4 \text{ years}$, $D \approx 2.6 \times 10^{-4} - 2.3 \times 10^{-3}$.

Although the effective duration is longer than the strict pericentre passage, the duty cycle remains low ($D \lesssim 0.6\%$). This confirms that the signals manifest as isolated transient events ("popcorn") rather than continuous sources.

\subsection{Detectability}

We estimate the detectability of individual eruptions using an illustrative $m=10\,M_{\odot}$ compact object at the Galactic Centre ($D=8.3$ kpc), orbiting the central MBH ($M=4.3\times 10^6\,M_{\odot}$). We assume a representative pericentre distance $R_p=10\,GM/c^2$. To fully characterize this system, we also define the orbital period $P=1$ year (the boundary case for $T_{\rm obs}=1$ year), yielding an eccentricity $e\approx 0.9974$. The eruption characteristics are determined using Keplerian approximations for the dynamics at closest approach.

The characteristic frequency of the eruption, $f_{\rm erupt}$, corresponds to the inverse timescale of the pericentre passage. Kinematically, this is estimated as:
\begin{equation}
    f_{\rm erupt} \approx \frac{1}{\pi}\sqrt{\frac{GM}{R_p^3}} = \frac{1}{10\sqrt{10}\pi} \frac{c^3}{GM}.
\end{equation}

\noindent Using the gravitational timescale $GM/c^3 \approx 21.18$ s, we find $f_{\rm erupt} \approx 0.475$ mHz.

The effective duration of the gravitational wave emission during the passage is estimated by the fly-by timescale:
\begin{equation}
    dt_{\rm peri} \approx \sqrt{\frac{R_p^3}{2GM}} = \sqrt{500} \frac{GM}{c^3} \approx 474 \, \text{s}.
\end{equation}

\noindent The peak strain amplitude $h$ is estimated using the leading-order quadrupole approximation for the emission at pericentre:
\begin{equation}
    h \approx \frac{2 G^2 M m}{c^4 D R_p} = \frac{1}{5} \frac{G m}{c^2 D} \approx 1.15\times 10^{-17}.
\end{equation}

\noindent A detailed calculation of the Signal-to-Noise Ratio (SNR) using the Peters formalism is presented in Section~\ref{sec:numerical_results}. For this illustrative system, we find an SNR $\approx 730$ for a $10\,M_{\odot}$ source and $\approx 2920$ for a $40\,M_{\odot}$ source. This confirms that individual eruptions from the Galactic Centre are detectable with very high confidence.

\section{Spectral Analysis of Transient Signals}
\label{sec:schwartz}

The analysis presented in this section addresses a critical requirement for the study of popcorn EMRIs: the definition of the Fourier transform for intrinsically transient signals. This foundational work is essential, not superfluous, because conventional spectral analysis techniques, optimized for continuous signals, fail to correctly capture the spectral content of isolated eruptions. Applying standard practices like windowing introduces artifacts and distortions that compromise the physical interpretation of the signal.

We analyze the mathematical framework required for the frequency domain representation of popcorn EMRI signals. These emissions manifest as short eruptions, potentially with discontinuities. Conventional spectral analysis techniques, developed primarily for continuous signals observed over finite times, must be applied with caution to intrinsically transient phenomena. We utilize the properties of the Schwartz functions to provide a basis for the Fourier transform of these eruptions and contrast it with the issues of spectral leakage inherent in standard data analysis practices.

\subsection{Spectral Leakage and Windowing in Conventional Analysis}

In gravitational wave data analysis, the Fourier transform (typically via the Fast Fourier Transform, FFT) is applied to finite segments of data \citep{Maggiore2008, ThraneTalbot2019}. The analysis of an infinitely long, continuous signal $h(t)$ over a finite duration $T$ is mathematically equivalent to observing a windowed signal $h_w(t)$:
\begin{equation}
    h_w(t) = h(t) \cdot \Pi_T(t),
\end{equation}

\noindent where $\Pi_T(t)$ is the rectangular window function (unity within the observation interval, zero otherwise).

By the convolution theorem, the observed spectrum $\tilde{h}_w(f)$ is the convolution of the true spectrum $\tilde{h}(f)$ and the Fourier transform of the window function:
\begin{equation}
    \label{eq:convolution_theorem}
    \tilde{h}_w(f) = \tilde{h}(f) * \mathcal{F}\{\Pi_T(t)\}.
\end{equation}

\noindent The Fourier transform of the rectangular window is the sinc function:
\begin{equation}
    \mathcal{F}\{\Pi_T(t)\} = T \, \mathrm{sinc}(\pi f T).
\end{equation}

\noindent The sinc function possesses significant side lobes that decay only as $1/f$. The convolution in Eq.~(\ref{eq:convolution_theorem}) smears the true spectrum; power from a spectral feature at frequency $f_0$ "leaks" into adjacent frequencies due to these side lobes. This phenomenon is spectral leakage.

In practice, the signal is also sampled at discrete times. This process is mathematically modeled by multiplying the continuous signal by a Dirac comb (Shah function), $\Sh_{\Delta t}(t) = \sum_{n=-\infty}^{\infty} \delta(t - n\Delta t)$, where $\Delta t$ is the sampling interval. The resulting analysis via the Discrete Fourier Transform (DFT) yields a spectrum characterized by the convolution with the Dirichlet kernel $D_N(\omega)$, which is the discrete-time Fourier transform of the rectangular window of length $N$ samples:
\begin{equation}
    D_N(\omega) = \frac{\sin(N\omega/2)}{\sin(\omega/2)}.
\end{equation}

\noindent The Dirichlet kernel exhibits the same problematic side lobe behavior as the sinc function.

Spectral leakage is primarily driven by the sharp discontinuities introduced by the rectangular window at the boundaries of the observation interval when analyzing continuous signals. If the signal amplitude is non-zero at these boundaries, the windowing process creates artificial high-frequency content associated with the abrupt truncation.

Standard mitigation involves applying a taper function $w(t)$ (e.g., Hann or Tukey) instead of the rectangular window. Tapers smoothly transition the signal to zero at the boundaries, reducing the discontinuities. The Fourier transforms of these taper functions have significantly suppressed side lobes compared to the Dirichlet kernel, thereby reducing long-range spectral leakage, albeit at the cost of broadening the main lobe (reducing frequency resolution).

\subsection{The Case of Popcorn EMRIs: Intrinsic Transients}

The situation with popcorn EMRIs differs fundamentally. The signal $h_{\rm erupt}(t)$ is intrinsically transient; it has significant amplitude only during the short pericentre passage $dt_{\rm peri}$ and is effectively zero otherwise. The morphology of the eruption, including its start and end (even if modeled as discontinuous), represents physical characteristics of the signal, not artifacts of the data acquisition length.

Applying a taper function to an intrinsic eruption signal fundamentally alters the data analysis problem. If a taper $w(t)$ is applied to the segment containing the eruption, the analyzed signal becomes:
\begin{equation}
    h_{\rm tapered}(t) = h_{\rm erupt}(t) \cdot w(t).
\end{equation}

\noindent The resulting spectrum is $\tilde{h}_{\rm tapered}(f) = \tilde{h}_{\rm erupt}(f) * \tilde{w}(f)$. This convolution with the taper's spectrum distorts the physical spectrum of the eruption.

More critically, tapering inherently reduces the signal power. By Plancherel's theorem (or Parseval's identity), the total energy is preserved under the Fourier transform. Since $|w(t)| \leq 1$ and the function is designed to approach zero at the boundaries, the total energy of the tapered signal is strictly less than the original signal (unless $w(t)=1$ everywhere, the rectangular window).
\begin{align}
    E_{\rm tapered} =& \int |h_{\rm tapered}(t)|^2 dt < \int |h_{\rm erupt}(t)|^2 dt = E_{\rm erupt}.
\end{align}

\noindent This reduction in signal energy directly impacts detectability. The optimal Signal-to-Noise Ratio (SNR) is given by the matched filtering integral. If the analysis relies on the tapered spectrum or if templates are not matched to the tapered signal, the calculated SNR will be systematically underestimated:
\begin{equation}
    \text{SNR}^2_{\rm tapered} < \text{SNR}^2_{\rm true}.
\end{equation}

\noindent This underestimation can be substantial. For a continuous signal, a Hann window reduces the signal power by a factor of 3/8 \citep{Harris1978}. We quantify this expected reduction for the $10\,M_{\odot}$ eruption analyzed in Section~\ref{sec:numerical_results}. Tapering the analysis segment with a Hann window reduces the expected signal power of a randomly placed transient by a factor of 3/8. Consequently, the root-mean-square expectation for the signal-to-noise ratio decreases from 730 to 447. The exact attenuation depends on the position of the eruption relative to the peak of the window function. For popcorn EMRIs, where the goal is to characterize the physical parameters encoded in the specific spectral shape of the eruption, such distortion and energy loss are unacceptable. It is therefore essential to establish why the direct Fourier transform (equivalent to a rectangular window perfectly encompassing the eruption) is the correct approach, rather than relying on ad hoc windowing practices borrowed from the analysis of continuous signals. We require a formalism that defines the Fourier transform of the intrinsic eruption signal without artificial modification.

\subsection{The Schwartz Space Formalism}

We analyze these signals using the density properties of Schwartz functions, which justifies the use of the direct Fourier transform (equivalent to using a rectangular window that fully encompasses the eruption) for transient signals.

The \textit{Schwartz space} $\mathscr{S}(\mathbb{R})$ is the space of all smooth, \textit{i.e.}, infinitely differentiable functions whose derivatives of any order vanish at infinity faster than every polynomial. More specifically a function $f:\mathbb{R}\to\mathbb{C}$ is a \textit{Schwartz function}, \textit{i.e.}, an element of $\mathscr{S}(\mathbb{R})$ if and only if
\begin{equation}
    \sup_{t\in\mathbb{R}}\lvert t^m f^{(n)}(t)\rvert < +\infty,
\end{equation}

\noindent for all non-negative integers $m,n$. A key property is that the Fourier transform operator $\mathcal{F}$ maps continuously $\mathscr{S}(\mathbb{R})$ onto itself.

Gravitational wave signals, particularly transients with compact support (which may be discontinuous), do not belong to $\mathscr{S}(\mathbb{R})$. However, they possess finite energy and duration, placing them within larger function spaces, specifically the spaces $L^p(\mathbb{R})$ of $p$-integrable functions, consisting of all complex-valued functions $f$ on $\mathbb{R}$ satisfying $\|f\|_{L^p}:=\left[\int_{\mathbb{R}}|f|^p\,dt\right]^{1/p}<+\infty$. For every such function $f$, its Fourier transform $\mathcal{F}f$ can be defined by a density argument \citep{Hormander1990ALPDO1}. The idea here is to construct a sequence $(f_n)$ of smooth functions that converge to $f\in L^p(\mathbb{R})$ in the norm $\|.\|_{L^p}$, and then define $\mathcal{F}f$ as the limit\footnote{Given that $\mathscr{S}(\mathbb{R})$ is a dense subspace of the space $\mathscr{S}'(\mathbb{R})$, this limit always exists and is generally a tempered distribution, not necessarily a function. However, the Fourier transforms of functions we are concerned with are well-defined either integrable or square-integrable, with the latter case corresponding to signals of finite energy.} of the sequence $\mathcal{F}f_n$. It is important here to note that, if $1\leq p\leq 2$, then $\mathcal{F}$ is a continuous operator mapping $L^p(\mathbb{R})$ to $L^q(\mathbb{R})$, where $q$ is the {H\"older conjugate} of $p$ satisfying $1/p + 1/q = 1$. This is a consequence of the Hausdorff-Young inequality.\footnote{See \cite{reed1972functional} Theorem IX.8.}

\section{Illustration: Analysis of a Rectangular Pulse}

We consider a transient pulse represented by a rectangular function:
\begin{equation}
    \label{eq:rect_pulse_def}
    f(t) = A \cdot \chi_{[-T/2, T/2]}(t),
\end{equation}

\noindent where $\chi_I(t)$ is the characteristic function of the interval $I$. This function is in $L^p(\mathbb{R})$ but not in $\mathscr{S}(\mathbb{R})$ since it's discontinuous at $t=\pm T/2$. We select the sequence of normalized Gaussian functions as the approximate identity, which is an element of the Schwartz space:
\begin{equation}
    \phi_n(t) = \sqrt{n/\pi} e^{-nt^2} \in \mathscr{S}(\mathbb{R}).
\end{equation}

\noindent The smooth approximating sequence $(f_n)$ is generated by convolution $f_n(t) = (f * \phi_n)(t)$. Explicitly:
\begin{align}
    f_n(t) =& \frac{A}{2} \left[ \mathrm{erf}\left(\sqrt{n}(t+T/2)\right) - \right. \left. \mathrm{erf}\left(\sqrt{n}(t-T/2)\right) \right].
\end{align}

\noindent We calculate $\mathcal{F}f_n(\omega)$ using the convolution theorem. Since $f\in L^1(\mathbb{R})$ and $\phi_n\in L^1(\mathbb{R})$,
\begin{equation}
    \mathcal{F}f_n(\omega) = \mathcal{F}f_{L^1}(\omega) \cdot \mathcal{F}\phi_n(\omega),
\end{equation}

\noindent where $\mathcal{F}f_{L^1}(\omega)$ denotes the standard Fourier integral of $f(t)$.
The Fourier transform of the Gaussian $\phi_n(t)$ is
\begin{equation}
    \mathcal{F}\phi_n(\omega) = e^{-\omega^2/(4n)}.
\end{equation}

\noindent The Fourier integral of the rectangular pulse $f(t)$ is
\begin{equation}
    \mathcal{F}f_{L^1}(\omega) = A T \, \mathrm{sinc}(\omega T/2).
\end{equation}

\noindent Thus, the Fourier transform of the smooth approximation is
\begin{equation}
    \mathcal{F}f_n(\omega) = e^{-\omega^2/(4n)} \cdot A T \, \mathrm{sinc}(\omega T/2).
\end{equation}

\noindent Notice that in the limit as $n\rightarrow\infty$, the sequence $(\mathcal{F}f_n)$ converges to $\mathcal{F}f_{L^1}(\omega)$. This demonstrates how the smoothing techniques presented here define the Fourier transform of $f(t)$ as the standard integral:
\begin{equation}
    \mathcal{F}f(\omega) = A T \, \mathrm{sinc}(\omega T/2).
\end{equation}

\section{Illustration: Analysis of a Truncated Sinusoid}

We analyze a transient signal modeled by a truncated sinusoidal wave:
\begin{equation}
    \label{eq:transient_sine_def}
    f(t) = A \sin(\omega_0 t) \cdot \chi_{[0, T]}(t).
\end{equation}

\noindent This function is bounded and has compact support, and is therefore in $L^p(\mathbb{R})$ for all $p\geq 1$. We construct the smooth approximating sequence $f_n(t) = (f * \phi_n)(t)$ using the Gaussian kernel $\phi_n \in \mathscr{S}(\mathbb{R})$. The Fourier transform of $f_n(t)$ is
\begin{equation}
    \label{eq:gaussian_fourier}
    \mathcal{F}f_n(\omega) = e^{-\omega^2/(4n)} \cdot \mathcal{F}f_{L^1}(\omega),
\end{equation}

\noindent where $\mathcal{F}f_{L^1}(\omega)$ is the standard Fourier integral:
\begin{align}
    \mathcal{F}f_{L^1}(\omega) = \frac{A}{2} \left( \frac{1-e^{i(\omega_0-\omega)T}}{\omega_0-\omega} + \frac{1-e^{-i(\omega_0+\omega)T}}{\omega_0+\omega} \right).
    \label{eq:fourier_transient_calc}
\end{align}

\noindent The limit $\lim_{n\rightarrow\infty}\mathcal{F}f_n(\omega)$ converges to Eq.~(\ref{eq:fourier_transient_calc}), validating the use of the standard Fourier integral for this transient signal.

\section{Numerical validation on a test signal}

We validate the Schwartz space formalism through numerical simulation of a transient gravitational wave test signal. We demonstrate the convergence of the smoothed approximations in both time and frequency domains.

\subsection{Simulated Signal Parameters}

We simulate a short gravitational wave strain signal $h(t)$, modeled as a truncated sinusoid (as defined in Eq.~\ref{eq:transient_sine_def}). The parameters are: amplitude $A = 10^{-21}$, characteristic frequency $f_0 = 50$ Hz ($\omega_0 = 100\pi$ rad/s), and duration $T = 0.5$ s. The strain $h(t)$ starts at $t_{\rm start}=0.5$ s and ends at $t_{\rm end}=1.0$ s.

\subsection{Numerical Smoothing and Convergence}

We construct the smooth approximation $h_n(t) = (h * \phi_n)(t)$. In the frequency domain, the smoothing acts as a low-pass filter (Eq.~\ref{eq:gaussian_fourier}). We require $4n \gg \omega_0^2 \approx 98696$. We select two values for the smoothing parameter: $n_1=50000$ and $n_2=500000$. For $n_1$, the attenuation factor at $\omega_0$ is $e^{-\omega_0^2/(4n_1)} \approx 0.61$. For $n_2$, the attenuation factor is $e^{-\omega_0^2/(4n_2)} \approx 0.95$. Figure~\ref{fig:time_domain_smoothing} shows the behavior of the smoothed signal $h_n(t)$ compared to the original strain data $h(t)$. The convolution process regularizes the discontinuities. As $n$ increases, $h_n(t)$ converges closely to $h(t)$.

\begin{figure}
    \centering
    \includegraphics[width=0.45\textwidth]{NormalizedStrain.pdf}
    \caption{Time domain smoothing of the transient test signal. The original discontinuous signal $h(t)$ (black dashed line) is approximated by smooth functions $h_n(t)$ for $n=5\times 10^4$ (red solid line) and $n=5\times 10^5$ (blue solid line). Convergence to $h(t)$ improves as $n$ increases.}
    \label{fig:time_domain_smoothing}
\end{figure}

\subsection{Spectral Analysis}

\begin{figure}
    \centering
    \includegraphics[width=0.45\textwidth]{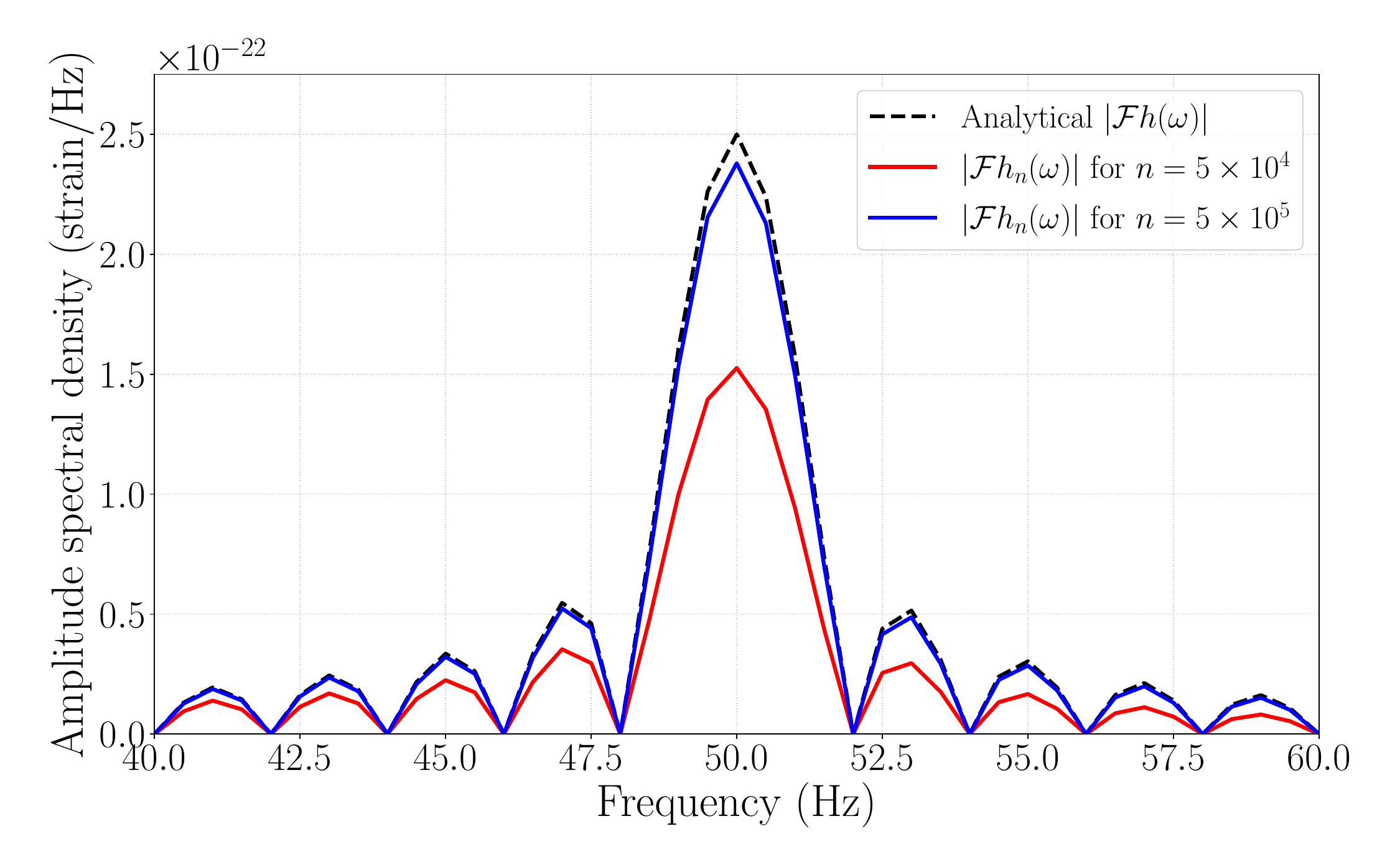}
    \caption{Frequency domain convergence of the Fourier transform magnitude. The analytical spectrum (black dashed line) is compared with the spectra of the smoothed signals for $n=5\times 10^4$ (red solid line) and $n=5\times 10^5$ (blue solid line). Attenuation caused by the smoothing decreases as $n$ increases.}
    \label{fig:frequency_domain_convergence}
\end{figure}

We compute the magnitude of the Fourier transform, $|\mathcal{F}h(\omega)|$. Figure~\ref{fig:frequency_domain_convergence} illustrates the convergence in the frequency domain. For $n_1=50000$, the spectrum shows attenuation near $f_0=50$ Hz due to the Gaussian filter effect. For $n_2=500000$, the spectrum closely matches the analytical spectrum, confirming the convergence discussed in Section~\ref{sec:schwartz}.

\section{Analysis of EMRI Eruptions in the Peters Approximation}

We apply the technique of smoothing by Schwartz functions to analyze the gravitational wave signal corresponding to a single pericentre passage of a highly eccentric EMRI, modeled using the Peters quadrupole approximation \citep{PetersMathews1963,Peters1964}.

\subsection{Time-Domain Signal Definition and Transformation}

We define the transient eruption signal $h(t)$ by isolating a single pericentre passage. The signal $h(t)$ has compact support and finite energy, thus $h(t) \in L^1(\mathbb{R}) \cap L^2(\mathbb{R})$. The technique of smoothing by Schwartz functions established in Section~\ref{sec:schwartz} guarantees that the Fourier transform $\mathcal{F}h(\omega)$ is defined and corresponds to the standard definition of the Fourier transform of integrable functions.

\subsection{Harmonic Content in the Peters Approximation}

We model the underlying orbital waveform using the quadrupole approximation. The power radiated in the $n$-th harmonic, $L_n$, is given by
\begin{equation}
    L_n = \frac{32}{5} \frac{G^4 M^3 m^2}{c^5 a^5} g(n,e),
\end{equation}

\noindent where $g(n,e)$ is defined in terms of Bessel functions $J_k(x)$:
\begin{align}
    g(n,e) = \frac{n^4}{32} &\left\{ \left[ J_{n-2}(ne) - 2eJ_{n-1}(ne) + \frac{2}{n}J_n(ne) \right. \right. \nonumber \\
    &+ \left. 2eJ_{n+1}(ne) - J_{n+2}(ne) \right]^2 \nonumber \\
    &+ (1-e^2)\left[ J_{n-2}(ne) - 2J_n(ne) + J_{n+2}(ne) \right]^2 \nonumber \\
    &+ \left. \frac{4}{3n^2} J_n(ne)^2 \right\}.
    \label{eq:g_n_e_full}
\end{align}

\subsection{Signal-to-Noise Ratio of the Eruption}

The optimal Signal-to-Noise Ratio (SNR) squared for the eruption is given by the matched filtering integral. This calculation relies on the existence of the Fourier transform $\tilde{h}(f) = \mathcal{F}h(2\pi f)$, which is guaranteed because $h(t) \in L^2(\mathbb{R})$, and defined via the Schwartz functions framework:
    \begin{equation}
    \label{eq:snr_integral_burst}
    \text{SNR}^2 = 4 \int_0^\infty \frac{|\tilde{h}(f)|^2}{S_n(f)} df.
\end{equation}

\noindent 
The Schwartz space formalism operates as a foundational argument. It shows that the untapered Fourier transform preserves the physical signal energy. This justifies computing the signal-to-noise ratio directly from the analytical radiated energy per harmonic, bypassing numerical time-domain windowing. We approximate the integral by a summation over the harmonics $f_n = n f_{\rm orb}$. This approximation connects the continuous energy spectrum of the eruption to the discrete energy emitted per harmonic during one orbit, $\Delta E_n = L_n P$. Averaged over orientation and polarization, the SNR is:
\begin{equation}
    \label{eq:snr_sum_burst}
    \text{SNR}^2 \approx \frac{2G}{\pi^2 c^3 D^2} \sum_{n=1}^{\infty} \frac{\Delta E_n}{f_n^2 S_n(f_n)}.
\end{equation}

\subsection{Numerical Results and Detectability}
\label{sec:numerical_results}

We apply this formalism to an illustrative popcorn EMRI system located at the Galactic Centre ($D=8.3$ kpc). We consider the central MBH mass $M=4.3\times 10^6 M_{\odot}$ and a compact object $m=10 M_{\odot}$. To satisfy the condition for a popcorn EMRI ($P>T_{\rm obs}$), we set the orbital period to $P=1$ year. We select a pericentre distance $R_p=10\,GM/c^2$. This represents an illustrative intermediate value in the strong-field regime.

These parameters define the system's geometry: the semi-major axis is $a\approx 3830\,GM/c^2$, and the eccentricity is $e\approx 0.9974$. The orbital frequency is $f_{\rm orb} \approx 31.7$ nHz. The characteristic frequency of the eruption is $f_{\rm erupt} \approx 0.475$ mHz.

For this highly eccentric system, the emission spectrum peaks at high harmonics. We calculate the cumulative SNR of the single eruption up to $N=100,000$ harmonics, $\text{SNR}(N)$, using Eq.~(\ref{eq:snr_sum_burst}), the exact expression for $g(n,e)$ in Eq.~(\ref{eq:g_n_e_full}), and a standard model for the LISA sensitivity $S_n(f)$. This truncation ensures that the neglected harmonics alter the cumulative signal-to-noise ratio by less than $10^{-4}$. We evaluate the high-order Bessel functions using uniform asymptotic expansions to ensure numerical stability. We utilize the analytical sensitivity curve defined by \cite{RobsonEtAl2019}, which incorporates the Galactic confusion noise, and we apply the sky-averaged and polarization-averaged detector response to provide a representative estimate. Computing the specific antenna pattern for the exact coordinates of the Galactic Centre introduces a factor of order unity that depends on the specific orbital phase of the LISA constellation at the time of the eruption.

First, if we were to use a simple characteristic approximation evaluated at $f_{\rm erupt}$ (typical for eruption searches), we would find $\text{SNR}_{\rm char} \approx 181$. This underestimates the detectability because the eruption has a broad spectrum that extends into frequencies where the LISA noise is lower.

Using the full Peters formalism and summing contributions from all relevant harmonics, the total accumulated SNR of the eruption is:
\begin{equation}
    SNR \approx 730.
\end{equation}

\noindent This demonstrates that a single pericentre passage from a long-period EMRI at the Galactic Centre is readily detectable by LISA. For a $40\,M_{\odot}$ object under the same conditions, the SNR is approximately 2920.

\begin{figure}
    \centering
    \includegraphics[width=\columnwidth]{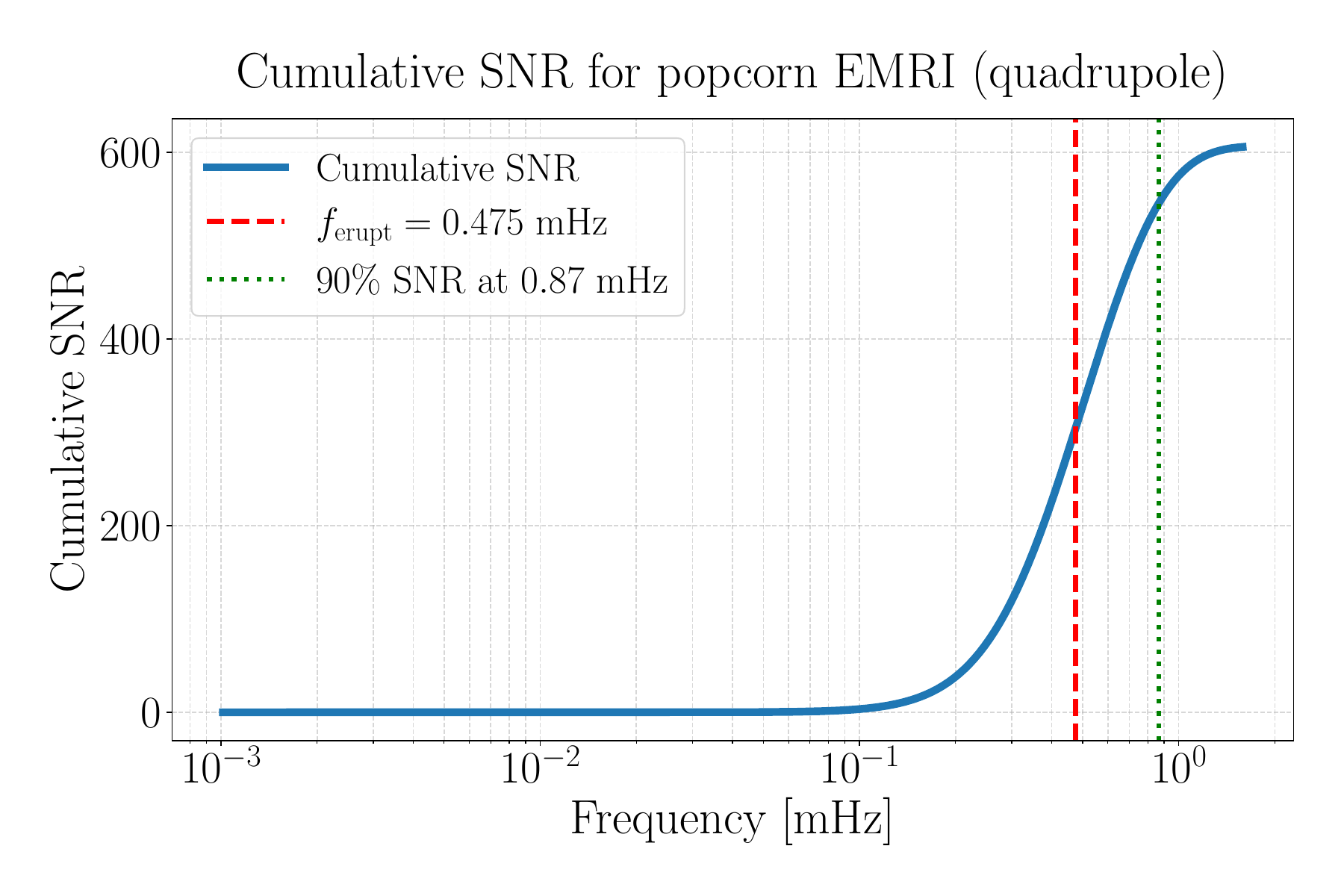}
    \caption{Cumulative SNR accumulation as a function of frequency for the illustrative popcorn EMRI system ($10 M_\odot$, $P=1$ yr, $e\approx 0.9974$) calculated using the Peters formalism. The characteristic frequency $f_{\rm erupt}$ (red dashed line) is located where the noise curve is steep. 90\% of the total SNR is accumulated by $f\approx 0.87$ mHz (green dotted line), where the noise level is significantly lower.}
    \label{fig:snr_accumulation}
\end{figure}

To ensure the robustness of these results, we performed a cross-validation using a time-domain inner product method. We computed the SNR by summing the signal power over the eruption duration, weighted by the noise variance derived from the instantaneous frequency $f_{\rm inst}(t)$. This approach, which does not rely on the Fourier decomposition into harmonics, yields consistent SNR values ($\approx 730$ for the $10\,M_{\odot}$ case), thereby validating the spectral analysis.

We note that the SNR calculation assumes a one year observation time. For shorter or longer segments, the SNR will be different. This depends on the evolutionary phase of the source. Unlike parabolic bursts, popcorn EMRIs are bound, coherent sources; they must eventually return to pericentre. The probability of detecting an eruption is roughly proportional to the duty cycle $D \lesssim 0.6\%$, but the coherent nature of the source implies that a non-detection in one window does not mean the source has vanished; it is merely in a quiescent phase of its orbit.

\section{Distinction between Schwartz Regularization and Windowing}

It is important to address the potential misunderstanding that regularizing the signal via convolution with a mollifier sequence $\phi_n(t)$ would be operationally redundant because, in the limit of large $n$, the sequence approximates a Dirac delta distribution. One could argue that this procedure is equivalent to computing the Fourier transform of the raw signal, which corresponds to the application of a rectangular window. While this mathematical equivalence holds in the asymptotic limit, this objection conflates the definition of the transform with the standard signal processing technique of tapering. The two approaches differ fundamentally in their mathematical structure and their impact on the signal-to-noise ratio.

Standard analysis typically mitigates spectral leakage by applying a window function $w(t)$, such as a ``Planck-taper''. This operation constitutes a pointwise multiplication in the time domain, $h_w(t) = h(t) \cdot w(t)$. By the convolution theorem, the resulting spectrum is the convolution of the intrinsic signal spectrum with the Fourier transform of the window,
\begin{equation}
    \label{eq:window_convolution}
    \tilde{h}_w(f) = \int_{-\infty}^{\infty} \tilde{h}(f') \tilde{w}(f - f') \, df'.
\end{equation}

\noindent Because any smooth window $w(t)$ designed to suppress leakage must vanish or decay at the boundaries of the analysis segment, this operation irreversibly modifies the spectral density. More critically, it reduces the total physical energy of the signal,
\begin{equation}
    \label{eq:energy_loss}
    \int_{-\infty}^{\infty} |h(t) w(t)|^2 \, dt < \int_{-\infty}^{\infty} |h(t)|^2 \, dt.
\end{equation}

\noindent For a transient signal with extended wings, such as the eccentric inspirals discussed here, this energy reduction directly diminishes the signal-to-noise ratio.

In contrast, the Schwartz formalism utilizes convolution in the time domain, $h_n(t) = (h * \phi_n)(t)$. This maps to a pointwise multiplication in the frequency domain,
\begin{equation}
    \label{eq:mollifier_product}
    \tilde{h}_n(f) = \tilde{h}(f) \cdot \tilde{\phi}_n(f).
\end{equation}

\noindent The fact that $\phi_n(t)$ approaches a delta function---and consequently $\tilde{\phi}_n(f)$ approaches unity---is not a trivial redundancy but the central objective of the proof. It demonstrates that the sequence of smooth functions converges strictly to the unmodified, raw Fourier transform of the transient. This provides the rigorous mathematical justification for rejecting the use of tapers that overlap the eruption. It proves that the discontinuities at the boundaries of a contained transient are physical features rather than numerical artifacts, and that the direct Fourier transform (equivalent to the rectangular window) is the unique estimator that preserves the full information content and energy of the eruption.

If data analysis employs a flat-top window, such as a Tukey taper, with a flat region that completely encompasses the eruption, the operation is mathematically equivalent to applying a rectangular window over the signal support. The Schwartz formalism presented here provides the theoretical justification for this procedure. It proves that the unmodified Fourier transform---equivalent to an untapered rectangular extraction over the eruption duration---is the unique estimator that preserves the total physical energy.

\section{Conclusions}

This study presents an analysis of the population, detectability, and description of popcorn EMRIs—transient gravitational wave signals generated by long-period, highly eccentric inspirals. We have estimated the population of EMRIs with orbital periods exceeding the observational timescale, based on an analytical steady-state model of the Galactic Centre.

Convolving the signal in time with the mollifier functions as multiplying the Fourier transform by a Gaussian window in frequency. This property provides an advantage over standard data analysis, which applies tapers in the time domain. Time-domain tapering of short eruptions forces the physical signal to zero at the edges, which decreases the signal-to-noise ratio and causes energy loss. The mollifier acts as a smooth low-pass filter that eliminates Gibbs ringing and retains the core energy. The limit as $n$ approaches infinity flattens the frequency window to unity, showing that the untapered Fourier transform conserves the total signal energy. This mathematical limit exemplifies the Heisenberg-Gabor limit. The time-domain mollifier width scales proportionally to $1/\sqrt{n}$ into a Dirac delta distribution, and the frequency bandwidth increases proportionally to $\sqrt{n}$. Measuring the onset and cessation of the eruptions requires this wide frequency range, because the power spreads across a broad spectrum. Standard windowing eliminates this distribution, whereas the convolution method guarantees its preservation.

Based on a conservative, illustrative two-component model and a 1-year observation baseline, the predicted steady-state population of popcorn EMRIs ranges from approximately 209 to 2003. This significantly exceeds the estimated continuous EMRI population (10 to 88). The analysis indicates a steady-state distribution $dN/dP \propto P^{-2/3}$, implying that the expected number of detected eruptions is approximately half the number of continuous sources. We forecast an observable eruption rate of 5 to 44 events per year (for $T_{\rm obs}=1$ year), decreasing to 2 to 18 per year (for $T_{\rm obs}=4$ years). The low duty cycle ($D \lesssim 0.6\%$) confirms their manifestation as isolated transients.

The proximity of the Galactic Centre suggests high detectability for individual eruptions. We calculate the SNR using a realistic LISA noise model (including confusion noise). Using the Peters formalism we obtain substantial SNRs (SNR $\approx 730$ for an illustrative system of $10\,M_{\odot}$). Moreover, eruptions, contrary to bursts, originate from a coherent source, which will therefore allow us to break parameter degeneracies.

We established the mathematical foundation for the frequency domain analysis of these transient signals using the theory of tempered distributions, and specifically the properties of Schwartz functions. This analysis is essential because standard data analysis practices, such as windowing, are designed for finite segments of continuous data and are inappropriate for intrinsically transient signals. The point of this rigorous approach is to ensure that the analysis correctly captures the signal properties without distortion. We have demonstrated mathematically that windowing distorts the physical morphology of the eruption and leads to significant underestimation of the SNR.

We distinguish the mathematical regularization of transient signals via Schwartz mollifiers from the standard practice of data windowing. While windowing relies on time-domain multiplication that irreversibly attenuates signal energy, our formalism demonstrates that the direct Fourier transform emerges as the unique, energy-preserving limit of a smooth regularization sequence. This result proves that the use of the rectangular window is not a numerical simplification, but the mathematically required choice to maximize the signal-to-noise ratio for popcorn EMRIs. The smoothing techniques implemented here confirm that the direct Fourier transform is the spectral representation of a transient event. This provides a solid foundation for data analysis strategies, dictating that tapering functions must be avoided when they overlap isolated popcorn EMRI eruptions, to prevent distortion of the waveform and biases in parameter estimation. This study establishes popcorn EMRIs as a potentially significant and detectable population of gravitational wave sources.

\section*{Data availability}

Any data used in this analysis are available on reasonable request
from the first author.




\bsp    
\label{lastpage}
\end{document}